\documentclass[twocolumn,pra,aps,superscriptaddress]{revtex4}
\usepackage{epsfig}
\usepackage{graphicx}
\usepackage{dcolumn}
\usepackage{bm} 
\usepackage{epstopdf}
\usepackage{amsmath}
\newcommand{\beq}{\begin{equation}}
\newcommand{\eeq}{\end{equation}}
\newcommand{\beqa}{\begin{eqnarray}}
\newcommand{\eeqa}{\end{eqnarray}}

\newcommand{\om}{\omega}

\def\jpb#1{{ J.\ Phys.\ B} {\bf#1}}

\def\pra#1{{ Phys.\ Rev. A\/} {\bf#1}}

\def\prl#1{{ Phys.\ Rev.\ Lett.} {\bf#1}}
\def\pr#1{{ Phys.\ Rev.\/} {\bf#1}}

\def\rmp#1{{ Rev.\ Mod.\ Phys.} {\bf#1}}

\begin{document}

\title{Time-domain theory of atomic photoionization}
\author{Rui-Hua Xu}
\affiliation{Graduate School, China Academy of Engineering Physics,  Beijing 100193, China}
\author{Zhaoyan Zhou}
\affiliation{Department of Physics, National University of Defense Technology, Changsha 410073, China}
\author{Xu Wang}
\email{xwang@gscaep.ac.cn}
\affiliation{Graduate School, China Academy of Engineering Physics,  Beijing 100193, China}

\date{\today}

\begin{abstract}
We present a combined numerical and theoretical study of atomic photoionization in the time domain. We show how a photoelectron wave packet rapidly changes its shape after being emitted, from a complex multi-peak structure to eventually a relatively simple single-peak structure. This time-domain shape evolution provides information beyond the time-dependent average position of the wave packet, which has been used to retrieve the Wigner time delay. For few-cycle laser pulses, the asymptotic velocity of the photoelectron can be different from long-pulse-based expectations due to non-negligible changes of the dipole matrix element within the spectra of the laser pulses.
\end{abstract}

\maketitle

\section{Introduction}

The photoelectric effect known today can be traced back to the experiments by H. Hertz \cite{Hertz1887} and by P. Lenard \cite{Lenard1902} in the late nineteenth and the early twentieth century. These experiments had motivated Einstein's quanta theory of light \cite{Einstein1905}. Existing in any material, the photoelectric effect has been extensively studied also in the context of atomic photoionization \cite{Cooper1962, Fano1968, Kennedy1972, Wuilleumier1974, Johnson1979, Starace1982, Yeh1985, Becker1989, Amusia1990, Chang1993, Becker1996}, especially with the advent of synchrotron radiations, which allow generation of high-energy photons with tunable frequencies.

Virtually all early studies on atomic photoionization, and on the photoelectric effect in general, focused on the energy or the momentum domain, mainly for the reason that the available light pulses were long (on the scale of picoseconds to nanoseconds or longer) and the time resolution was not sufficient for the photoionization process. With rapid advancements of ultrafast laser technologies, however, light pulses with femtosecond or sub-femtosecond durations can be routinely generated. Table-top titanium:sapphire lasers can generate near-infrared pulses of femtosecond durations. Free electron lasers can generate femtosecond pulses in the X-ray regime. Based on high harmonic generation, isolated light pulses of durations of a few tens of attoseconds have been reported \cite{Zhao2012, Li2017, Gaumnitz2017}. Hard X-ray pulses of 200-attoseconds duration have also been reported using free-electron lasers at the Linac Coherent Light Source \cite{Huang2017}.

These ultrashort light sources have enabled investigations of ultrafast dynamical processes in the time domain, such as following electron localization in molecules\cite{Sansone2010}, real-time observation of Auger decay \cite{Drescher2002}, real-time observation of electron tunneling ionization\cite{Uiberacker2007}, charge transfer in polyatomic molecules \cite{Erk2016, Rudenko2017}, real-time buildup of Fano resonances \cite{Kaldun2016}, attosecond time delays in atomic photoionization \cite{Schultze2010, Pazourek2015, Kheifets2010, Moore2011, Nagele2012, Dahlstrom2012, Dahlstrom2012jpb, Kheifets2013, Feist2014, Maquet2014, Wei2016}, etc.

The availability of these ultrashort light sources makes it desirable to (re-)think about the photoionization process specially from the time-domain perspective. In fact, the work by Eisenbud \cite{Eisenbud1948}, Wigner \cite{Wigner1955}, and Smith \cite{Smith1960} discussed time-domain aspects of photoionization before the appearance of any ultrashort light sources. The so-called Wigner (or Eisenbud-Wigner-Smith) delay is time-domain information extracted from the average position ($\langle r \rangle (t)$, i.e., the first moment) of the emitted photoelectron wave packet \cite{Pazourek2015}. The goal of the current article is to go beyond the average positions. By numerically solving a time-dependent Schr\"odinger equation (TDSE) in full dimensions, we show how the {\it shape} of the photoelectron wave packet evolves with time, which supplies richer information than the average positions. We find that the photoelectron wave packet has a rather complex multi-peak structure right after being emitted, but it rapidly adjusts its shape, reduces the number of peaks, and eventually evolves into a relatively simple single-peak structure. This time-dependent shape evolution process will be explained from the temporal evolution of the phase of the photoelectron wave packet.

We will also show that for few-cycle laser pulses, the asymptotic velocity of the photoelectron can be different from normal, long-pulse-based expectations. For short laser pulses with relatively broad spectra, the dipole transition matrix element may change appreciably within the spectra of the laser pulses, shifting the average momentum and the asymptotic velocity of the photoelectron wave packet. Depending on the detailed form of the dipole transition matrix element, the photoelectron wave packet may be faster or slower than that generated with long pulses.

This article is organized as follows. In Section II we introduce the methods that we use, including a numerical method for solving the TDSE and the time-dependent perturbation theory for interpretation of numerical results. In Section III we present our numerical results and the corresponding analyses and discussions. A conclusion is given in Section IV.

\section{Methods}

\subsection{Numerical solution of the time-dependent Schr\"odinger equation}

The TDSE for atomic hydrogen interacting with an external laser field can be written as (in atomic units)
\begin{equation}
i\frac{\partial}{\partial t}\psi(\mathbf{r},t)=\hat{H}\psi(\mathbf{r},t)=[\hat{H}_{0}+\hat{H}_{I}]\psi(\mathbf{r},t),
\end{equation}
where $\hat{H}_{0}$ is the field-free Hamiltonian and $\hat{H}_{I}$ is the atom-field interaction
\begin{eqnarray}
\hat{H}_{0} &=& -\frac{1}{2}\frac{d^{2}}{dr^2}+\frac{\hat{L}^2}{2r^2}+V(r),\\
\hat{H}_{I} &=& \mathbf{r} \cdot \mathbf{\hat{e}_z} \varepsilon(t) = \varepsilon(t) r \cos \theta. \label{e.H_I}
\end{eqnarray}
For the hydrogen atom, $V(r)=-1/r$ and we have used the length-gauge form of the interaction Hamiltonian. The laser field $\varepsilon(t)= \varepsilon_{0} f(t) \cos\omega t$ is assumed to be linearly polarized along the $z$ direction with amplitude $\varepsilon_{0}$ and angular frequency $\omega$. A trapezoidal pulse envelope function $f(t)$ has been used, which has a two-cycle turning on and a two-cycle turning off
\begin{eqnarray}
f(t)=\left\{\begin{array}{ll}
t/2T  & \ \ \textrm{$0< t \le 2T$}\\
1  & \ \ \textrm{$2T < t \le \tau-2T$}\\
(-t+\tau)/2T & \ \ \textrm{$\tau-2T < t \le \tau$}
\end{array}\right.
\end{eqnarray}
where $T=2\pi/\omega$ is the duration of an optical cycle and $\tau$ is the duration of the whole pulse.

We use a generalized pseudospectral method \cite{Tong1997} to numerically solve the TDSE.
The Schr\"odinger equation can be propagated in discrete time steps as
\begin{eqnarray}
\psi(\mathbf{r},t+\Delta t)&\simeq&\exp(-i\hat{H}_{0}\Delta t/2 )\nonumber\\
&\times& \exp[-i\hat{H}_{I}(r,\theta,t+\Delta t)\Delta t]\nonumber\\
&\times& \exp(-i\hat{H}_{0}\Delta t/2)\psi(\mathbf{r},t)+O(\Delta t^3)
\label{psiyanhua}
\end{eqnarray}
The time propagation of the wave function from $t$ to $t+\Delta t$ is achieved by three steps: (i) Propagation for half a time step $\Delta t/2$ in the energy space spanned by $\hat{H}_{0}$; (ii) Transformation to the coordinate space and propagation for one time step $\Delta t$ under the atom-field interaction $\hat{H}_{I}$; (iii) Transformation back to the energy space spanned by $\hat{H}_{0}$ and propagation for another half time step $\Delta t/2$. The commutation errors are on the order of $\Delta t^3$.

The wave function $\psi(\mathbf{r},t)$ can be expanded in Legendre polynomials
\begin{eqnarray}
\psi(r_{i},\theta_{j},t)=\sum_{l=0}^{lmax}g_{l}(r_{i})P_{l}(\cos\theta_{j}),
\end{eqnarray}
if the atom is initially in an $s$-state (the magnetic quantum number $m=0$) and the laser polarization is linear ($\Delta m=0$). The $g_{l}(r_{i})$ is calculated by the Gauss-Legendre quadrature
\begin{eqnarray}
g_{l}(r_{i})=\sum_{k=1}^{L+1}w_{k}P_{l}(\cos\theta_{k})\psi(r_{i},\theta_{k},t),
\end{eqnarray}
where quadrature lattices ${\cos\theta_{k}}$ are zeros of the Legendre polynomials $P_{L+1}(\cos\theta_{k})$ and ${w_{k}}$ is the corresponding quadrature wights.

Now the evolution of the wave function in the energy space spanned by $\hat{H}_{0}$  can be written as
\begin{align}
\exp(&-i\hat{H}_{0}\Delta t/2)\psi(r_{i},\theta_{j},t) \nonumber \\
&=\sum_{l=0}^{lmax}[\exp(-i\hat{H}_{0}^l\Delta t/2)g_{l}(r_{i},t)]P_{l}(\cos\theta_{j}).
\label{eq3.23}
\end{align}
Each $g_{l}$ is propagated independently within individual $\hat{H}_{0}^{l}$ energy space.

In order to avoid artificial boundary reflection, for each time step a mask function $M(r)=\cos^{1/4}[\frac{\pi}{2}(r-r_{0})/(r_{m}-r_{0})]$ is multiplied to the wave function for $r\ge r_0$. Here $r_{0}$ is the entrance radius of the absorbing region and $r_{m}$ is the radius of the numerical grid.

\subsection{The time-dependent perturbation theory}

\begin{figure}
  \centering
  \includegraphics[width=3.8cm, height=4.5cm]{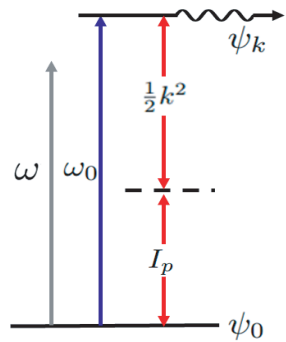}
  \caption{Schematic illustration of the energies (frequencies) involved in atomic photoionization. The laser frequency is $\om$, which induces a transition from a bound state $\psi_0$ to a possible continuum state $\psi_k$. The energy difference between the two states is $\om_0$, which may not be in exact resonance to $\om$. $\om_0$ can be approximately divided into the ionization potential $I_p$ and the kinetic energy of the emitted electron $k^2/2$, if the ponderomotive shifts can be neglected. }\label{f.illustration}
\end{figure}

\begin{figure*}
  \centering
  \includegraphics[width=16cm, trim=0 0 -20 0]{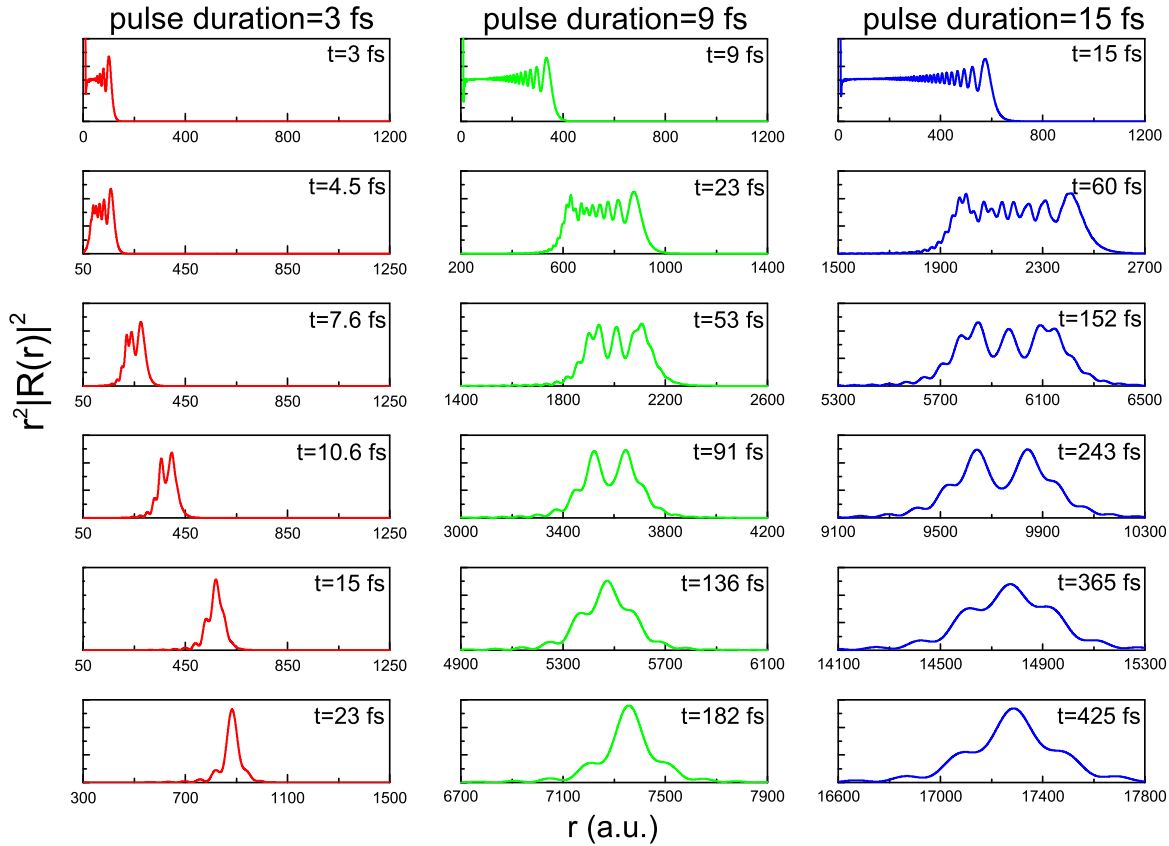}\\
  \caption{Time evolution of the shape of the emitted photoelectron wave packet for three different laser pulse durations, namely, 3 fs (left column), 9 fs (middle column), and 15 fs (right column). The time for each snapshot is given in the upper right corner of each panel. For all the three cases, the laser frequency $\om = 1.0$ a.u., corresponding to 0.15 fs per optical cycle, and the laser peak intensity is $10^{13}$ W/cm$^2$.
  }\label{f.wpshape}
\end{figure*}

From the first-order perturbation theory, the transition amplitude from an initial (bound) state $\psi_0 (\mathbf{r})$ to a final continuum state $\psi_k (\mathbf{r})$ is
\beqa
c(k,t) &=& -i\int_{0}^{t} dt' \langle \psi_k(\mathbf{r}) | \hat{H}_I(t) | \psi_0(\mathbf{r}) \rangle e^{i\om_0 t'} \nonumber \\
&=& -i \varepsilon_0 D(k) \int_{0}^{t} dt' \cos \om t' e^{i \om_0 t'}
\eeqa
where $D(k)\equiv \langle \psi_k (\mathbf{r}) | r\cos\theta |  \psi_0 (\mathbf{r}) \rangle$ is the dipole transition matrix element, $\om_0$ is the energy difference between the two states, and for the purpose of analytical simplicity we have assumed $f(t)=1$. $\psi_0({\bf r})$ and $\psi_k(\mathbf{r})$ are eigenstates of the unperturbed Hamiltonian $\hat{H}_0$ and they can be obtained numerically by solving the corresponding eigenvalue problem. We can proceed
\beqa
c(k,t) &=& -i\frac{\varepsilon_0 D(k)}{2} \int_{0}^{t} dt' \left[ e^{i(\om_0 + \om)t'} + e^{i(\om_0 - \om)t'} \right] \nonumber \\
&=& -\frac{\varepsilon_0 D(k)}{2} \left[ \frac{e^{i(\om_0+\om)t}-1}{\om_0+\om}
 + \frac{e^{i(\om_0-\om)t}-1}{\om_0-\om}  \right] \nonumber\\
& &
\eeqa

For laser frequencies not far away from resonance, i.e., $|\om_0 - \om| \ll \om_0 + \om$, we can apply the rotating wave approximation by neglecting the first term in the square braket
\beqa
c(k,t) &\approx& -\frac{\varepsilon_0 D(k)}{2} \frac{e^{i(\om_0-\om)t}-1}{\om_0-\om}   \nonumber \\
&=& -i \varepsilon_0 D(k) \frac{\sin[(\om_0-\om)t/2]}{\om_0-\om} e^{i(\om_0-\om)t/2} \label{e.ckt}
\eeqa
In the above formula, $\om_0 = \om_0 (k) = I_p + k^2/2$ (for weak laser fields where ponderomotive shifts can be neglected), as illustrated schematically in Fig. \ref{f.illustration}.

The ionized wave packet for a particular partial wave in the configuration space is given by the radial wave function
\beqa
R_l(r,t) = \int_{0}^{\infty} dk \ c(k,t) R_{kl} (r) e^{-ik^2t/2}. \label{e.Rrt}
\eeqa

\section{Results and discussions}

\subsection{Shape evolution of the photoelectron wave packet}

\begin{figure*}
  \centering
  \includegraphics[width=5.2cm]{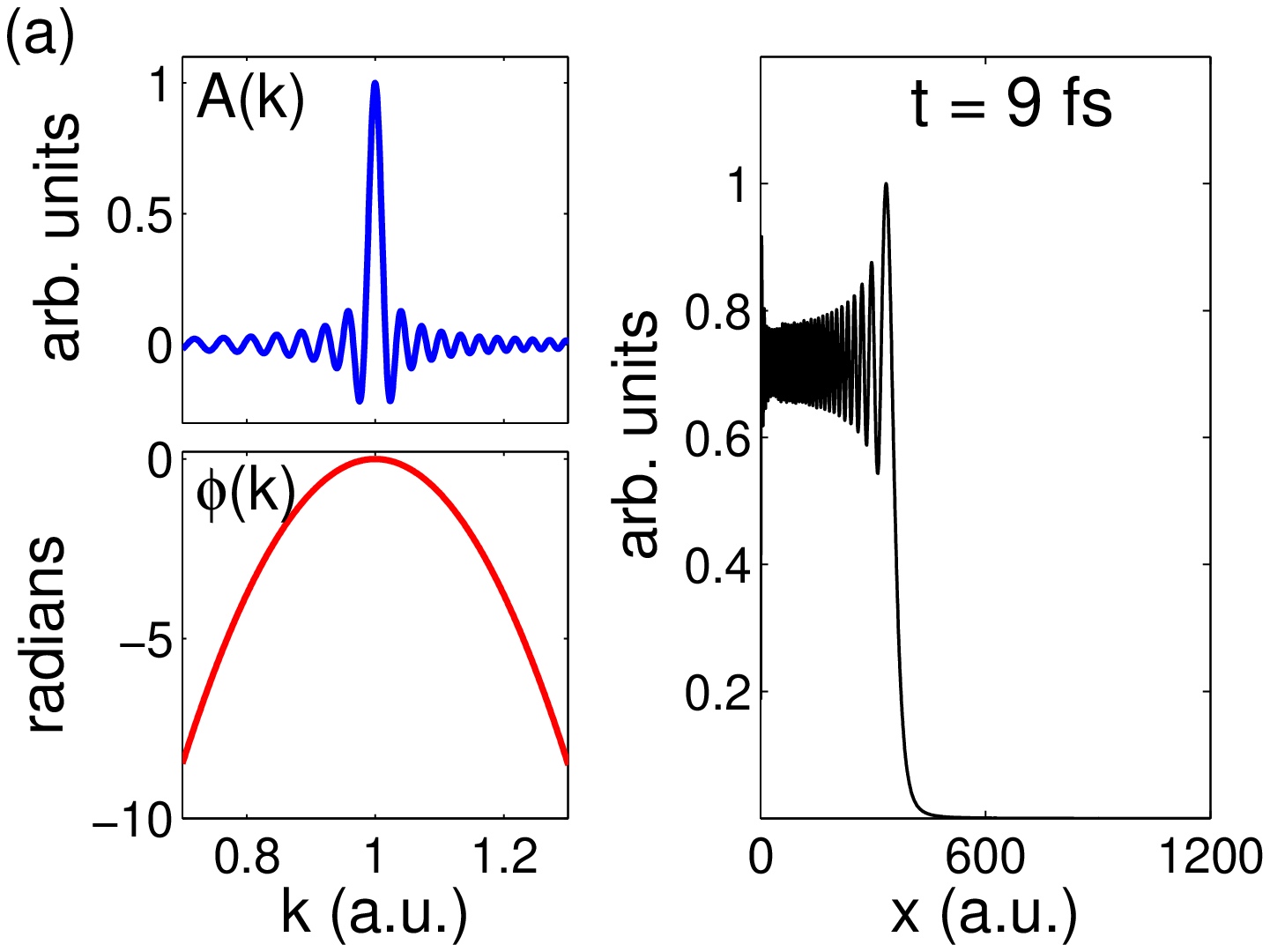}
  \hspace{0.3cm}
  \includegraphics[width=5.2cm]{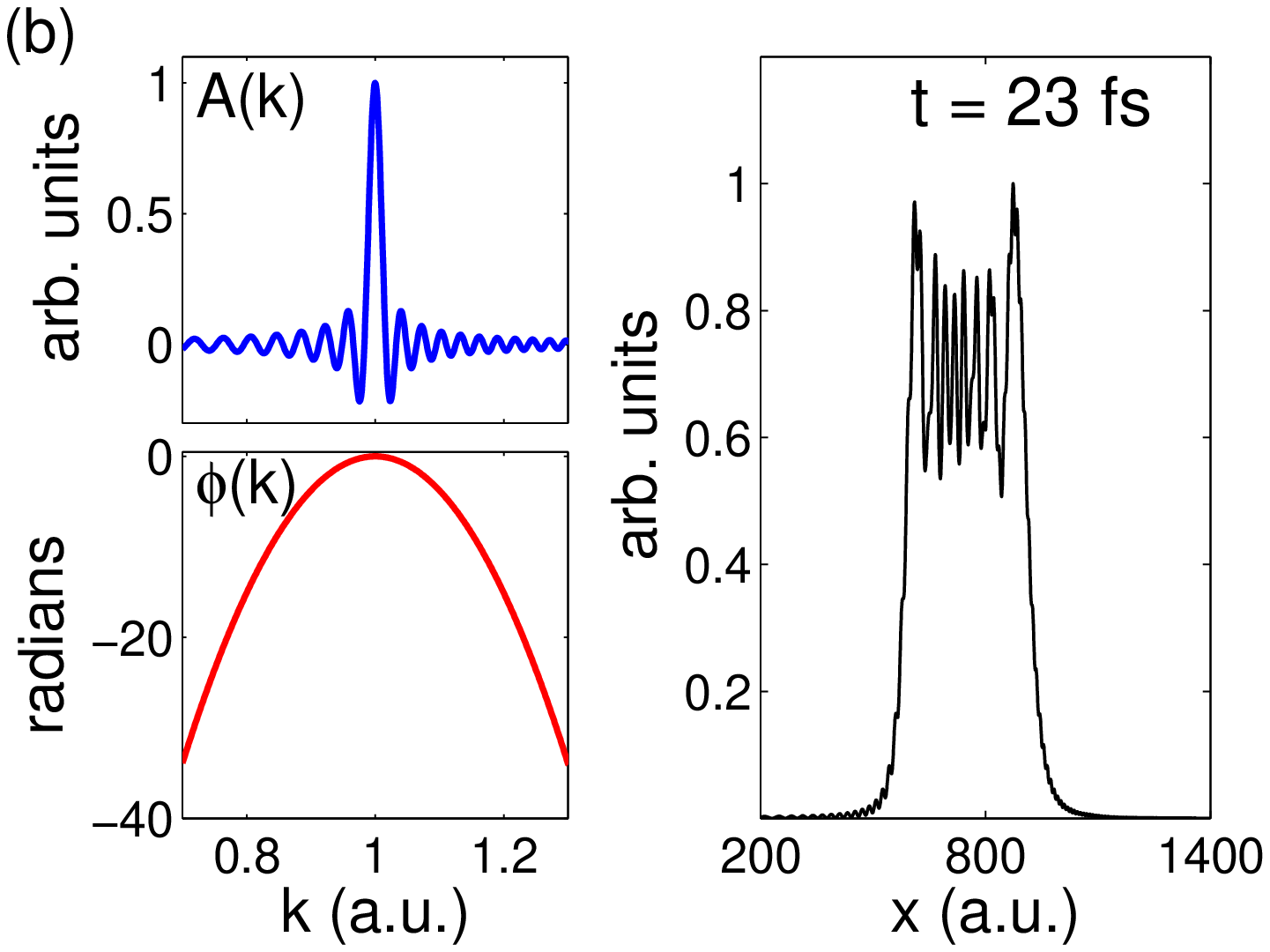}
  \hspace{0.3cm}
  \includegraphics[width=5.2cm]{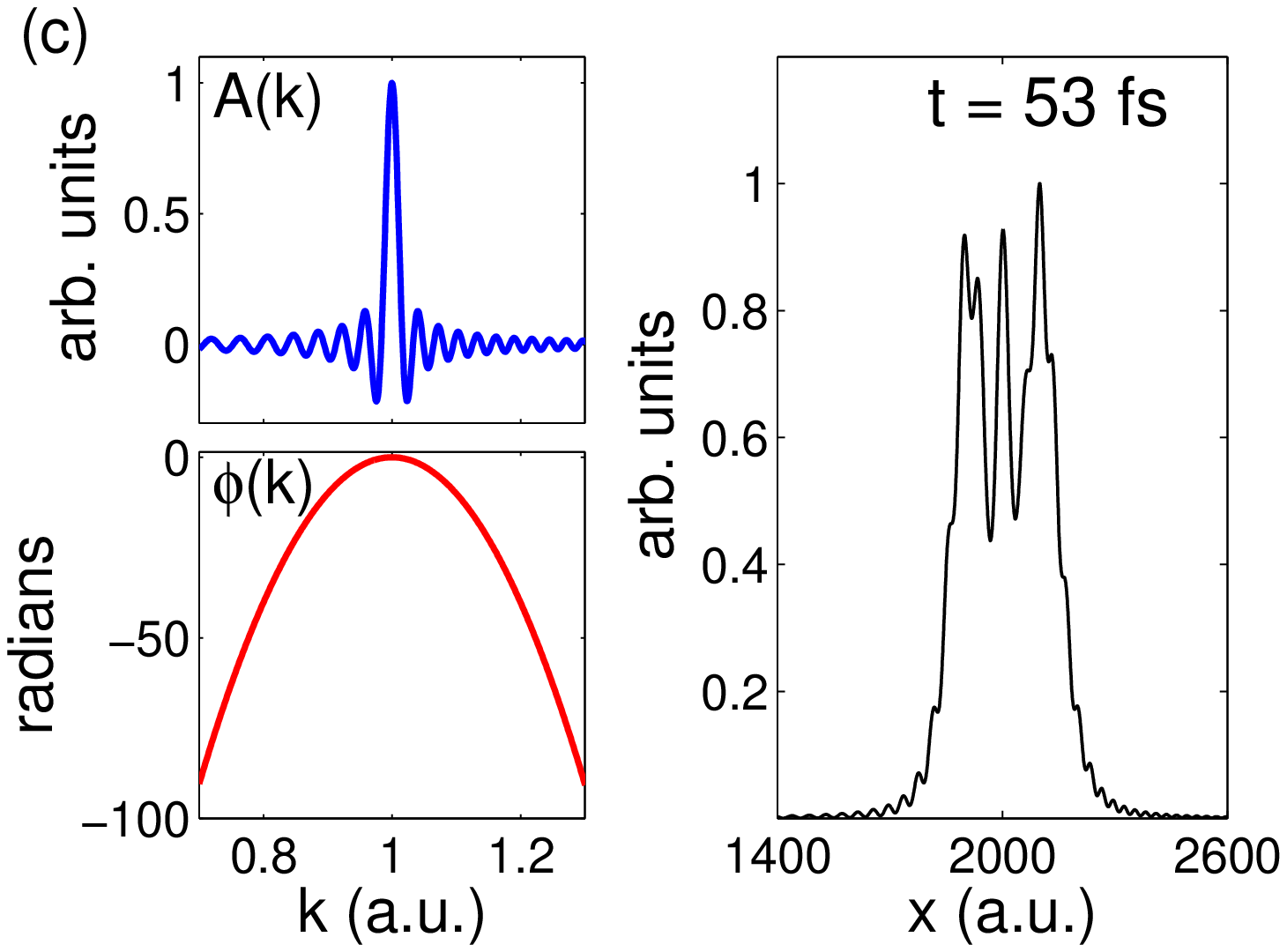}\\
  \vspace{0.3cm}
  \includegraphics[width=5.2cm]{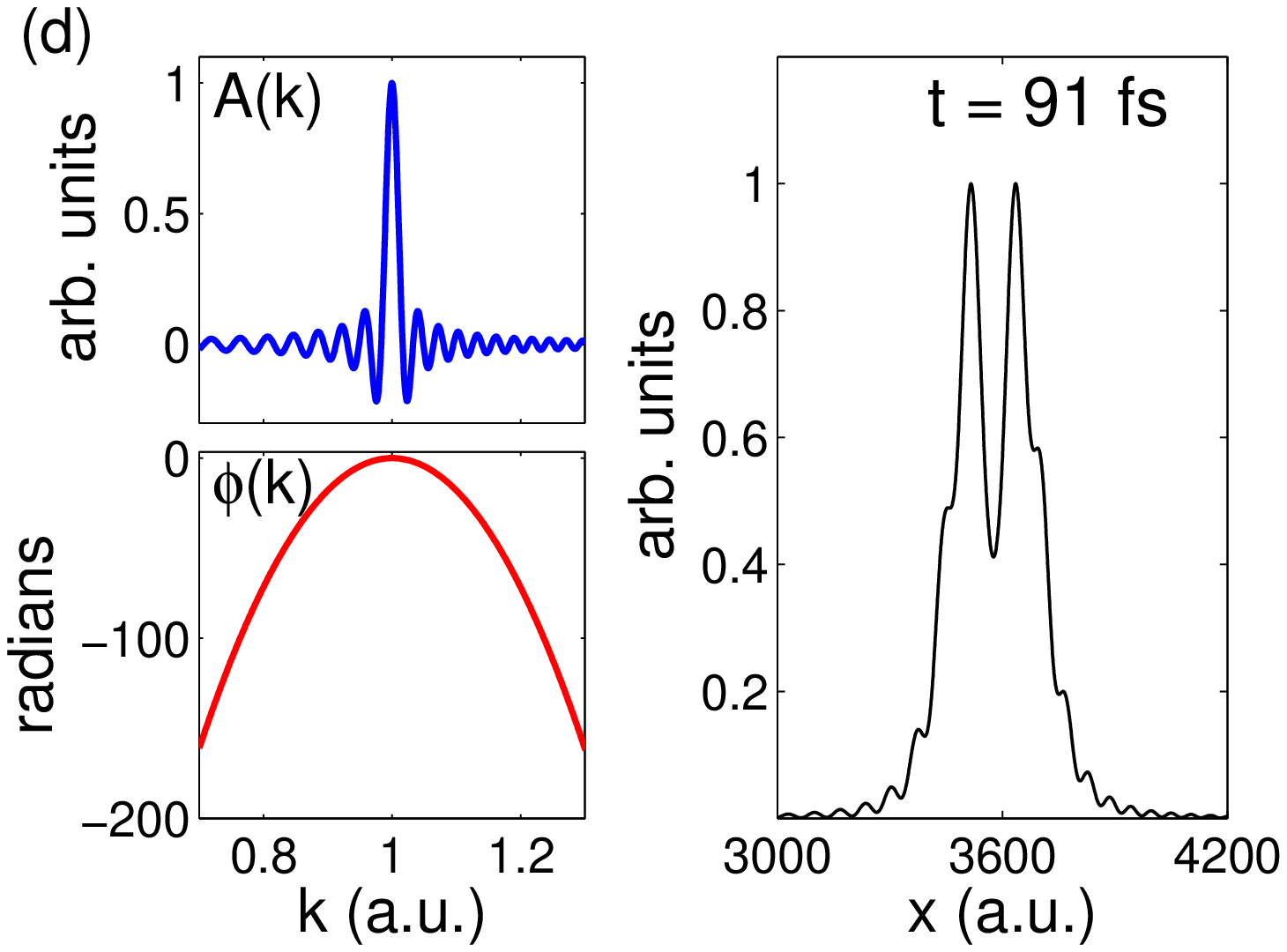}
  \hspace{0.3cm}
  \includegraphics[width=5.2cm]{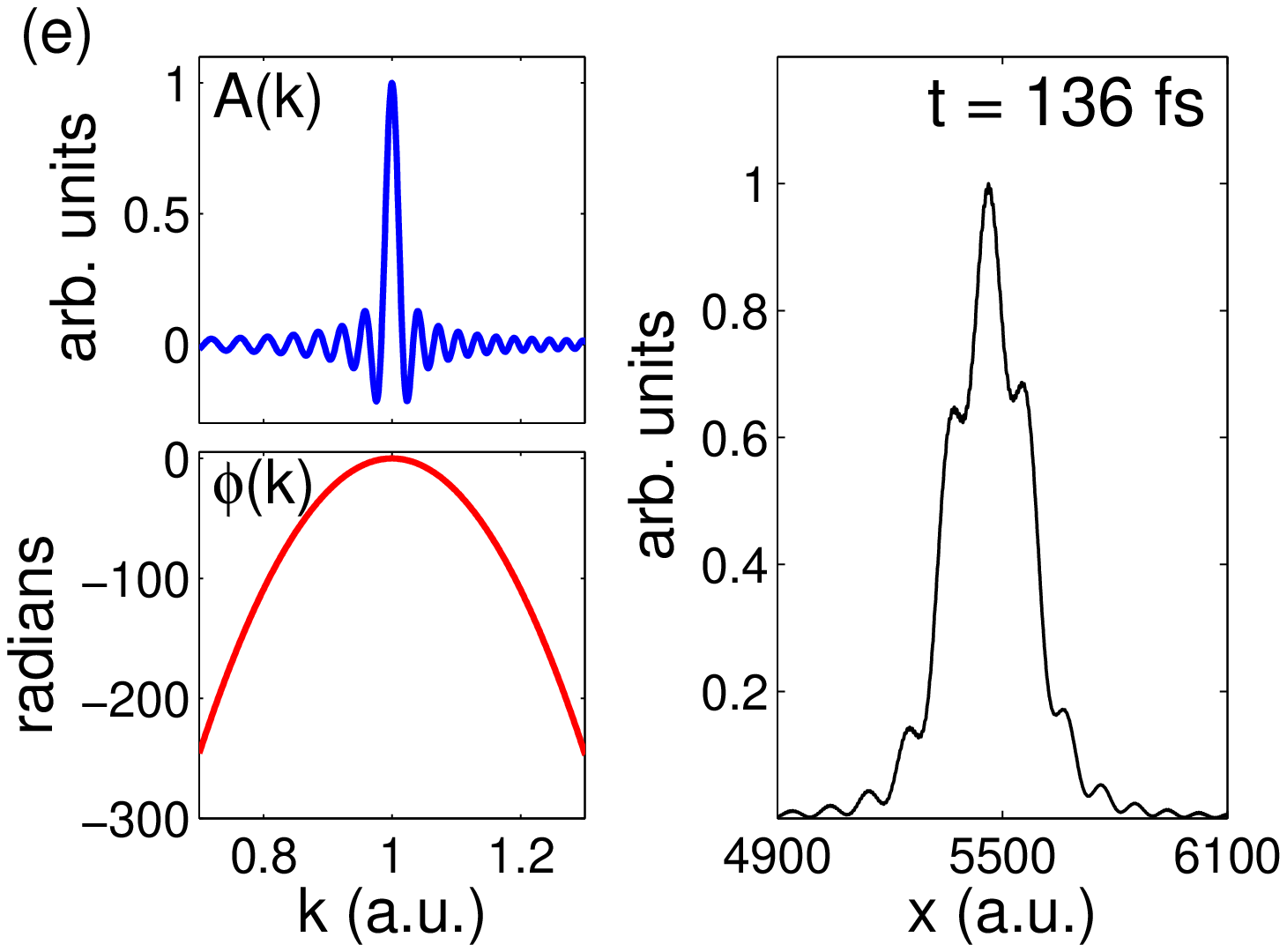}
  \hspace{0.3cm}
  \includegraphics[width=5.2cm]{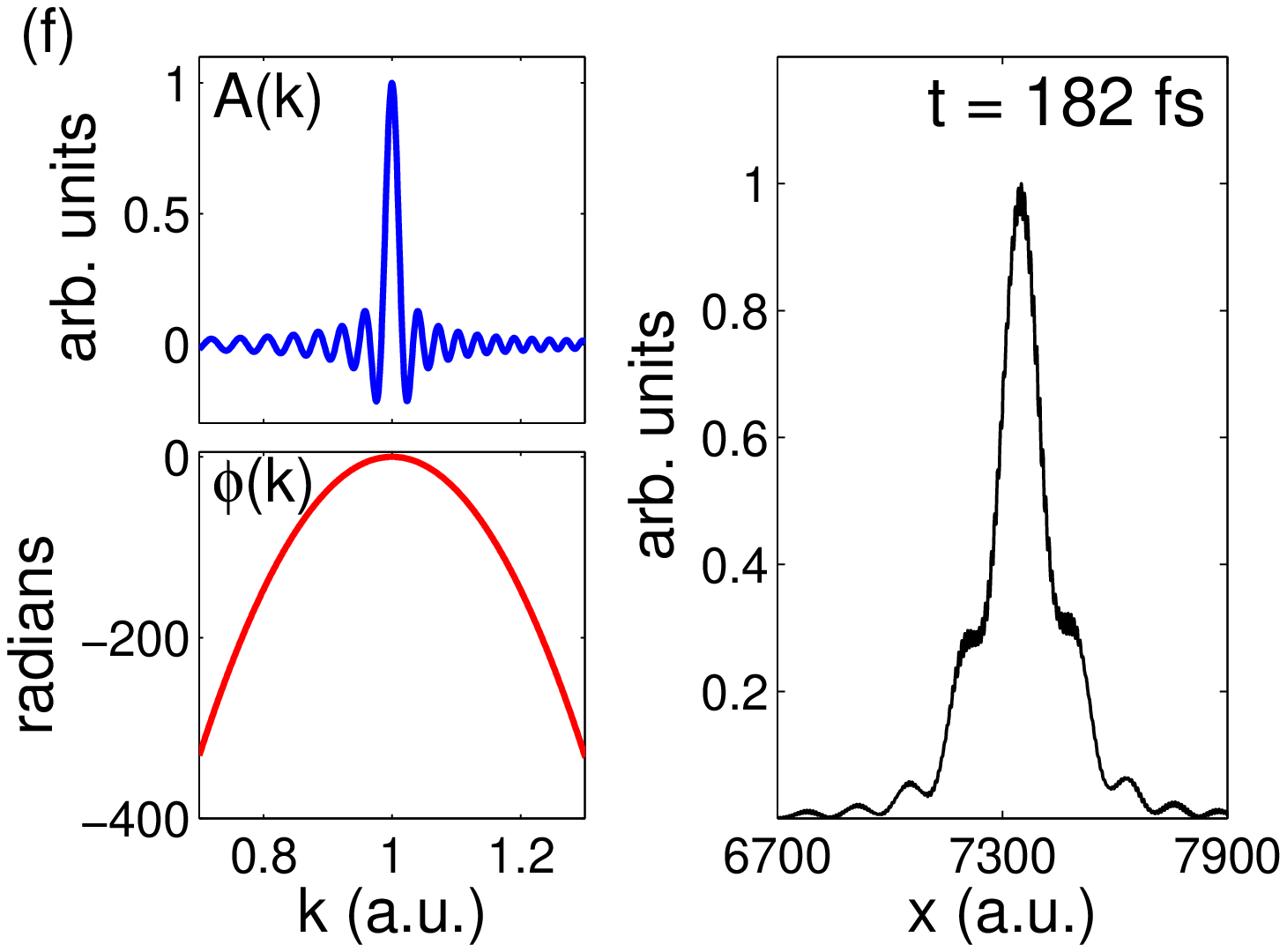}
  \caption{The function $A(k)$, the quadratic component of the function $\phi(k,t)$, and the resulting shapes of the photoelectron wave packet predicted by Eq. (\ref{e.psixtmodel}). Notice the vertical scale of $\phi(k)$ for each panel. The shapes of the wave packets are to be compared with the middle column of Fig. \ref{f.wpshape}.}\label{f.modelshape}
\end{figure*}

Figure \ref{f.wpshape} shows time-domain shape evolutions of photoelectron wave packets, for three different laser pulse durations, namely, 3 fs (left column), 9 fs (middle column), and 15 fs (right column). The laser intensity $10^{13}$ W/cm$^2$ and the laser frequency is $\om = 1.0$ a.u. For exact resonance where $\om_0 = \om$, the wave number is denoted as $k_0$, which satisfies
\beq
\frac{k_0^2}{2} = \om - I_p = 0.5 \ \text{a.u.}, \ \text{or} \ \ k_0 = 1\ \text{a.u.} \label{e.k0}
\eeq
For each panel, the horizontal axis is the distance (in atomic units) from the remaining ion, and the vertical axis is the radial wave packet $r^2|R(r)|^2$. Since we start from the ground state of the hydrogen atom, and the laser field is linearly polarized, only one partial wave ($l=1$, $m=0$) is involved in the continuum states.

We see from all the three examples that right after, or shortly after the laser pulse is over (e.g., the top two rows), the photoelectron wave packet has a rather complex multi-peak structure. As time evolves, the wave packet quickly adjusts its shape (while at the same time translating in space), and the number of peaks reduces. Eventually a relatively simple single-peak structure emerges. The time scale of this shape evolution process depends on the duration of the laser pulse. For the three pulse durations that we use in Fig. \ref{f.wpshape}, this shape adjusting process lasts for a few tens of femtoseconds to a few hundred femtoseconds. The wave packet will keep spreading in later times (not shown), but it will remain a single-peak structure.

This shape evolution of the photoelectron wave packet provides richer information beyond the average position $\langle r \rangle (t)$, extrapolation of which backward in time yields the Wigner time delay (see Ref. \cite{Pazourek2015} for a recent review). It remains open the possibility of utilizing the additional information (e.g., $\langle r^2 \rangle (t)$, $\langle r^3 \rangle (t)$, etc.) to facilitate understandings about the photoionization process.

\subsection{Qualitative understandings from the time-dependent perturbation theory}

The shape evolutions of the photoelectron wave packets given in Fig. \ref{f.wpshape} are obtained numerically by solving the full dimensional TDSE. In this subsection we will give a qualitative understanding of the main features using the time-dependent perturbation theory.

For simplicity we consider in one dimension and assume that the continuum states are plane waves $\psi_k(x) = e^{ikx}$. From Eqs. (\ref{e.ckt}) and (\ref{e.Rrt}) the wave packet in the configuration space is
\beqa
\psi(x,t) &=& -i \varepsilon_0 \int_{0}^{\infty} dk \ D(k) \frac{\sin \left[ \left( k^2 -k_0^2\right)\tau/4 \right]}{\left( k^2 - k_0^2\right)/2} \nonumber \\
& & \times \exp\left[ i \left( k^2 -k_0^2 \right) \tau/4 - i k^2 t/2 \right] e^{ikx} \label{e.psixtpw}
\eeqa
for $t\ge\tau$. Here we have referred Eq. (\ref{e.k0}) for the definition of $k_0$. Note that $c(k,t) = 1$ for $t>\tau$ but the wave packet still gains a free evolution phase $\exp(-ik^2t/2)$ after the pulse is over.

If the range of accessible continuum states is narrow, then the dipole matrix element $D(k)$ may be regarded as a constant within the range and moved out of the integral, replaced by value $D(k_0)$. This approximation is valid for relatively long pulses. For short few-cycle pulses, this approximation may break down and an example will be given in the following subsection.

Then the shape of the photoelectron wave packet can be written as
\beqa
\psi(x,t) &\approx& -i\varepsilon_0 D(k_0) \int_{0}^{\infty} dk \frac{\sin \left[ \left( k^2 -k_0^2\right)\tau/4 \right]}{\left( k^2 - k_0^2\right)/2} \nonumber \\
& & \times \exp\left[ i \left( k^2 -k_0^2 \right) \tau/4 - i k^2 t/2 \right] e^{ikx} \nonumber \\
&\equiv& -i\varepsilon_0 D(k_0) \int_{0}^{\infty} dk A(k) e^{i\phi(k,t)} e^{ikx} \label{e.psixtmodel}
\eeqa
where we have defined
\beqa
A(k) &=& \frac{\sin \left[ \left( k^2 -k_0^2\right)\tau/4 \right]}{\left( k^2 -k_0^2 \right)/2} \label{e.Ak}\\
\phi(k,t) &=& \left( k^2 - k_0^2 \right)\frac{\tau}{4} - \frac{k^2t}{2} \nonumber \\
&=& \frac{1}{2} \left( \frac{\tau}{2}-t \right) (k-k_0)^2 \nonumber\\
& &+ \left( \frac{\tau}{2}-t \right) k_0 (k-k_0) - \frac{t}{2} k_0^2 \label{e.phikt}
\eeqa
$A(k)e^{i\phi(k,t)}$ gives the weight of each plane wave component. We see that the function $\phi(k,t)$ contains a quadratic term centered on $k_0$, which is responsible for the shape of the photoelectron wave packet, a linear term, which is responsible for the spatial translation of the wave packet, and a constant term, which does not have a physical effect.

Fig. \ref{f.modelshape} shows the function $A(k)$, {\it the quadratic component of} $\phi(k,t)$, and the resulting shape of the photoelectron wave packet for six different times to be compared to the middle column of Fig. \ref{f.wpshape}. One sees pretty good qualitative agreements with the wave packet shapes obtained using the full dimensional TDSE, for all the six times. The wave packet starts from a complex multi-peak structure, then adjusts itself by reducing the number of peaks, eventually becomes a relatively simple single-peak structure.

\subsection{Photoionization in the few-cycle limit}

\begin{figure}
  \centering
  \includegraphics[width=8cm]{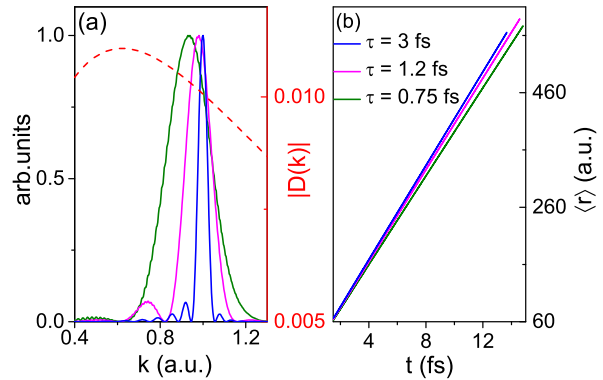}\\
  \caption{(a) Photoelectron momentum distributions obtained with three laser pulse durations, namely, 3 fs (blue), 1.2 fs (magenta), and 0.75 fs (green). The red dashed curve shows the shape of $|D(k)|$, the amplitude of the dipole transition matrix element (right vertical scale). (b) The corresponding time-dependent average position $\langle r \rangle (t)$ showing different asymptotic slopes.}\label{f.fewcycle}
\end{figure}

For very short laser pulses, e.g., pulses consisting only a few cycles, the dipole transition matrix element may change appreciably within the spectra of the laser pulses. Then in Eq. (\ref{e.psixtpw}) we cannot move $D(k)$ out of the integral by assigning its value to be $D(k_0)$. The consequence is that each plane wave component will be further weighted by $D(k)$, resulting in a shift in the photoelectron momentum distribution and a different asymptotic photoelectron velocity.

Figure \ref{f.fewcycle} shows the momentum distribution of the photoelectron for three different laser pulse durations, namely, 3 fs (blue), 1.2 fs (magenta), and 0.75 fs (green). As the pulse duration decreases, the photoelectron momentum distribution has two changes. First, the width increases; And second, the peak position shifts. In this example, the peak shifts to the slower side, that is, a red shift. This can be understood from the shape of the dipole transition matrix element, the amplitude of which is shown in the same panel as the red dashed curve. $|D(k)|$ decreases across the spectrum of the laser pulse (which is peaked in the neighborhood of $k=1$), so it effectively weights more the continuum states with smaller $k$ values. The consequence is a red shift in the photoelectron momentum distribution as the pulse duration decreases.

In the time domain the photoelectron will have different asymptotic velocities, as shown in the right panel of Fig. \ref{f.fewcycle}. The time-dependent average position of the photoelectron $\langle r \rangle (t)$ has different asymptotic slopes for the three laser pulses.

Whether a few-cycle laser pulse leads to a slower or a faster asymptotic photoelectron velocity depends on the shape of the dipole transition matrix element $D(k)$. If the amplitude of $D(k)$ increases across the spectrum of the laser pulse, then it puts more weights on the faster components, and the asymptotic velocity of the photoelectron will be faster than that obtained with longer laser pulses.

\section{Conclusion}

In this article we present a combined numerical and theoretical study of atomic photoionization in the time domain. Numerical results are obtained by solving the full dimensional time-dependent Schr\"odinger equation. And the numerical results are analyzed using the time-dependent perturbation theory. A time-domain study of the atomic photoionization process is motivated by rapid recent progresses of ultrafast laser technology, which can generate laser pulses as short as a few tens of attoseconds.

Time-domain aspects of atomic photoionization had been considered by Eisenbud, Wigner, and Smith. The so-called Wigner time delay can be retrieved from extrapolating the time-dependent average position ($\langle r \rangle (t)$, i.e., the first moment) of the photoelectron wave packet. This article goes beyond the average position by studying the time evolution of the shape of the photoelectron wave packet. We find that the shape of the photoelectron wave packet experiences rapid and dramatic changes after being emitted. Right after emission, the wave packet has a rather complex and multi-peak structure, but it evolves rapidly with time by reducing the number of peaks. Eventually, the wave packet has a rather simple single-peak structure to be detected by a detector.

This shape evolution process can be qualitatively explained by a simple model based on the time-dependent perturbation theory. The key factor is the quadratic component of the phase of the wave packet.

For long laser pulses, the spectra are usually narrow enough such that the dipole transition matrix element changes little within the spectra and can be treated as a constant. For short few-cycle laser pulses with relatively wide spectra, however, the dipole transition matrix element may change appreciably within the spectra of the laser. The consequence is a change in the photoelectron asymptotic velocity, either faster or slower, depending on the detailed form of the dipole transition matrix element.

We acknowledge funding support from China NSF No. 11774323, China Science Challenge Project No. TZ2018005, NSAF No. U1730449, and National Key R\&D Program No. 2017YFA0403200.


\begin{thebibliography}{*}

\bibitem{Hertz1887} H. Hertz, Ann. Phys. \textbf{267}, 983 (1887).

\bibitem{Lenard1902} P. Lenard, Ann. Phys. \textbf{313}, 149 (1902).

\bibitem{Einstein1905} A. Einstein, Ann. Phys. \textbf{322}, 132 (1905).

\bibitem{Cooper1962} J. W. Cooper, \pr{128}, 681 (1962).

\bibitem{Fano1968} U. Fano and J. W. Cooper, Rev. Mod. Phys. \textbf{40}, 441 (1968).

\bibitem{Kennedy1972} D. J. Kennedy and S. T. Manson, \pra{5}, 227 (1972).

\bibitem{Wuilleumier1974} F. Wuilleumier and M. O. Krause, \pra{10}, 242 (1974).

\bibitem{Johnson1979} W. R. Johnson and C. D. Lin, \pra{20}, 964 (1979).

\bibitem{Starace1982} A. F. Starace, in {\it Handbuch der Physik}, Vol XXXI, pp. 1-121, edited by W. Mehlhorn, Springer, Berlin (1982).

\bibitem{Yeh1985} J. J. Yeh and I. Lindau, Atomic Data and Nuclear Data Tables, {\bf 32}, 1 (1985).

\bibitem{Becker1989} U. Becker, D. Szostak, H. G. Kerkhoff, M. Kupsch, B. Langer, R. Wehlitz, A. Yagishita, and T. Hayaishi, \pra{39}, 3902 (1989).

\bibitem{Amusia1990} M. Ya. Amusia, {\it Atomic Photoeffect}, Plenum, New York (1990).

\bibitem{Chang1993} T. N. Chang (Ed.), {\it Many-Body Theory of Atomic Structure and Photoionization}, World Scientific, Singapore (1993).

\bibitem{Becker1996} U. Becker and D. A. Shirley (Ed.), {\it VUV and Soft X-Ray Photoionization Studies}, Plenum, New York (1996).

\bibitem{Zhao2012} K. Zhao, Q. Zhang, M. Chini, Y.  Wu, X. Wang, and Z. Chang, Opt. Lett. \textbf{37}, 3891 (2012).

\bibitem{Li2017} J. Li, {\it el al.}, Nat. Commun. \textbf{8}, 186 (2017).

\bibitem{Gaumnitz2017} T. Gaumnitz, A. Jain, Y. Pertot, M. Huppert, I. Jordan, F. Ardana-Lamas, and H. Jakob W\"{o}rner, Opt. Express \textbf{25}, 27506 (2017).

\bibitem{Huang2017} S. Huang, Y. Ding, Y. Feng, E. Hemsing, Z. Huang, J. Krzywinski, A. A. Lutman, A. Marinelli, T. J. Maxwell, and D. Zhu, \prl{119}, 154801 (2017).

\bibitem{Sansone2010}G. Sansone {\it el al.}, Nature \textbf{465},763 (2010).

\bibitem{Drescher2002} M. Drescher, M. Hentschel, R. Kienberger, M. Uiberacker, V. Yakovlev, A. Scrinzi, Th. Westerwalbesloh, U. Kleineberg, U. Heinzmann, and F. Krausz , Nature \textbf{419},803(2002).

\bibitem{Uiberacker2007} M. Uiberacker {\it et al.}, Nature \textbf{446},627(2007).

\bibitem{Erk2016} B. Erk {\it et al.}, Science {\bf 345}, 288 (2014).

\bibitem{Rudenko2017} A. Rudenko {\it et al.}, Nature {\bf 546}, 129 (2017).

\bibitem{Kaldun2016} A. Kaldun {\it et al.}, Science {\bf 354}, 738 (2016).

\bibitem{Schultze2010} M. Schultze {\it et al.}, Science \textbf{328}, 1658 (2010).

\bibitem{Pazourek2015} R. Pazourek, S. Nagele, and J. Burgd\"orfer, \rmp{87}, 765 (2015).

\bibitem{Kheifets2010} A. S. Kheifets and I. A. Ivanov, \prl{105}, 233002 (2010).

\bibitem{Moore2011} L. R. Moore, M. A. Lysaght, J. S. Parker, H. W. van der Hart, and K. T. Taylor, \pra{84}, 061404 (2011).

\bibitem{Nagele2012} S. Nagele, R. Pazourek, J. Feist, and J. Burgdorfer, \pra{85}, 033401 (2012).

\bibitem{Dahlstrom2012} J. M. Dahlstr\"om, T. Carette, and E. Lindroth, \pra{86}, 061402 (2012).

\bibitem{Dahlstrom2012jpb} J. M. Dahlstr\"om, A. L'Huillier, and A. Maquet, \jpb{45}, 183001 (2012).

\bibitem{Kheifets2013} A. S. Kheifets, \pra{87}, 063404 (2013).

\bibitem{Feist2014} J. Feist, O. Zatsarinny, S. Nagele, R. Pazourek, J. Burgdorfer, X. Guan, K. Bartschat, and B. I. Schneider, \pra{89}, 033417 (2014).

\bibitem{Maquet2014} A. Maquet, J. Caillat, and R. Ta\"ieb, \jpb{47}, 204004 (2014).

\bibitem{Wei2016} H. Wei, T. Morishita, and C. D. Lin, \pra{93}, 053412 (2016).

\bibitem{Eisenbud1948} L. Eisenbud, Ph.D. thesis, Princeton University, 1948.

\bibitem{Wigner1955} E. P. Wigner, Phys. Rev. {\bf 98}, 145 (1955).

\bibitem{Smith1960} F. T. Smith, Phys. Rev. {\bf 118}, 349 (1960).

\bibitem{Tong1997} X.-M. Tong and S.-I Chu, Chem. Phys. {\bf 217}, 119 (1997).






\end{thebibliography}
\end{document}